\documentclass[%
reprint,
superscriptaddress,
showpacs,
amsmath,amssymb,
aps,
prc,
twocolumn,
floatfix, ]%
{revtex4}

\usepackage{color}
\usepackage{CJK}
\usepackage{graphicx}
\usepackage{dcolumn}
\usepackage{bm}
\usepackage{epstopdf}
\usepackage{multirow}
\usepackage{booktabs}
\usepackage{longtable}
\usepackage[
bookmarks=true,colorlinks,%
citecolor=blue,linkcolor=blue,anchorcolor=blue,filecolor=blue,urlcolor=blue,%
]{hyperref}

\allowdisplaybreaks

\begin{document}

\begin{CJK*}{UTF8}{}
\CJKfamily{gbsn}
\title{Microscopic study of higher-order deformation effects
 on the ground states of superheavy nuclei around $^{270}$Hs}

\author{Xiao-Qian Wang (王晓倩)}
\affiliation{CAS Key Laboratory of Theoretical Physics, Institute of Theoretical Physics, Chinese Academy of Science, Beijing 100190, China}
\affiliation{School of Physical Sciences, University of Chinese Academy of Sciences, Beijing 100049, China}

\author{Xiang-Xiang Sun (孙向向)}
\email{sunxiangxiang@ucas.ac.cn}
\affiliation{School of Nuclear Science and Technology, University of Chinese Academy of Sciences, Beijing 100049, China}

\author{Shan-Gui Zhou (周善贵)}
\affiliation{CAS Key Laboratory of Theoretical Physics, Institute of Theoretical Physics, Chinese Academy of Science, Beijing 100190, China}
\affiliation{School of Physical Sciences, University of Chinese Academy of Sciences, Beijing 100049, China}
\affiliation{School of Nuclear Science and Technology, University of Chinese Academy of Sciences, Beijing 100049, China}
\date{\today}

\begin{abstract}
We study the effects of higher-order deformations $\beta_\lambda$ ($\lambda=4,6,8,$ and $10$)
on the ground state properties of superheavy nuclei (SHN)
near the doubly magic deformed nucleus $^{270}$Hs
by using the multidimensionally-constrained (MDC) relativistic mean-field (RMF) model
with five effective interactions PC-PK1, PK1, NL3$^{*}$, DD-ME2, and PKDD.
The doubly magic properties of $^{270}$Hs are featured
by the large energy gaps at $N=162$ and $Z=108$
in the single-particle spectra.
By investigating the binding energies and single-particle levels of $^{270}$Hs
in multidimensional deformation space,
we find that the deformation $\beta_6$
has the greatest impact on the binding energy among these higher-order deformations
and influences the shell gaps considerably.
Similar conclusions hold for other SHN near $^{270}$Hs.
Our calculations demonstrate that the deformation $\beta_6$ must be considered when studying SHN by using MDC-RMF.
\end{abstract}

\maketitle
\end{CJK*}

\section{Introduction}
\label{Sec:1}
One of the challenges in modern nuclear physics is exploring
the mass and charge limits of atomic nuclei
\cite{Hamilton2013_ARNPS63-383,
Nazarewicz2018_NatPhys14-537,
Giuliani2019_RMP91-011001,
Zhou2014_Phy43-817,
Li2014_NPR31-253,
Zhou2017_NPR34-318,
Zhou2019_Phys48-640,
Zhou2020_SciSinPMA50-112002_E}.
The prediction of the existence of an ``island of stability"
of superheavy nuclei (SHN) was made
\cite{Myers1966_NP81-1,
Wong1966_PL21-688,
Sobiczewski1966_PL22-500,
Meldner1967_ArkivF36-593,
Mosel1969_ZPA222-261,
Nilsson1969_NPA131-1}
in the 1960s.
The elements with $Z \leq 118$ have been synthesized
\cite{Hofmann2000_RMP72-733,Morita2015_NPA944-30,
Oganessian2017_PS92-023003} up to now.
Various predictions of the center of the ``island of stability" have been made
\cite{Myers1966_NP81-1,Wong1966_PL21-688,Sobiczewski1966_PL22-500,
Meldner1967_ArkivF36-593,Mosel1969_ZPA222-261,Nilsson1969_NPA131-1,
Rutz1997_PRC56-238,Zhang2005_NPA753-106,Sobiczewski2007_PPNP58-292}
and the position of this island is still not well determined.
Contrary to the ``island of stability",
the existence of a ``shallow" of SHN has been well established
theoretically and experimentally,
which connects the continent of stable nuclei to the ``island of stability" of SHN.
The center of this shallow is predicted to be around $Z = 108$ and $N = 162$ and
consists of deformed SHN
\cite{Moller1974_NPA229-292,Cwiok1983_NPA410-254,
Patyk1991_NPA533-132,Smolanczuk1995_PRC52-1871}.
$^{270}_{108}$Hs$_{162}$ is a doubly magic deformed nucleus
\cite{Dvorak2006_PRL97-242501,Oganessian2013_PRC87-034605} and
offers a prototype to explore the structure of SHN.

Nowadays, there are two kinds of theoretical approaches to study
structures and properties of SHN,
macroscopic-microscopic method (MMM) and microscopic method.
Generally, the surface of a nucleus is parameterized as
\cite{Ring1980},
\begin{equation}
R(\theta, \varphi)=
R_{0}\left[
1+\beta_{00}+
\sum_{\lambda=1}^{\infty}
\sum_{\mu=-\lambda}^{\lambda}
\beta_{\lambda \mu}^{*}
Y_{\lambda \mu}(\theta, \varphi)\right],
\end{equation}
where $\beta_{\lambda\mu}$ is deformation parameter and
$R_0$ is the radius of the sphere with the same volume.
There is a remarkable question,
how large should the dimension of the deformation space be
when studying deformed SHN?
In 1991, Patyk and Sobiczewski studied the ground state properties
of the heaviest even-even nuclei with proton numbers $Z = 90$--$114$
and neutron numbers $N = 136$--$168$ by using the MMM and
found that the $\beta_6$ degree of freedom is important for binding energies
and the formation of deformed shells
\cite{Patyk1991_NPA533-132,Patyk1991_PLB256-307}.
In addition,
$\beta_6$ also has a considerable influence on the moments of inertia
\cite{Muntian2001_PLB500-241,Liu2012_PRC86-011301R}
and the high-$K$ isomers
\cite{Liu2011_PRC83-011303R,He2019_ChinPC43-064106}.
The microscopic description of the structure for SHN can be
achieved by using density functional theories and
there are few works
investigating the influence of $\beta_6$ on the binding energy
and shell structure of SHN so far.

Covariant density functional theory (CDFT) is one of
the most successful self-consistent approaches
and has been used to describe ground states and
excited states of nuclei throughout the nuclear chart
\cite{Ring1996_PPNP37-193,
Bender2003_RMP75-121,Vretenar2005_PR409-101,Meng2006_PPNP57-470,
Paar2007_RPP70-691,Niksic2011_PPNP66-519,
Liang2015_PR570-1,Meng2015_JPG42-093101,
Zhou2016_PS91-063008}.
To study the ground state properties, potential energy surfaces (PESs),
and fission barriers of
heavy nuclei and SHN,
the multidimensionally-constrained (MDC) CDFTs
have been developed
\cite{Lu2012_PRC85-011301R,Lu2014_PRC89-014323,
Zhou2016_PS91-063008,Zhao2016_PRC93-044315}.
MDC-CDFTs have been applied to study hypernuclei
\cite{Lu2011_PRC84-014328,Lu2014_PRC89-044307,
Rong2020_PLB807-135533,Rong2021_arXiv2103.10706},
the fission barriers and the PESs of actinide nuclei
\cite{Lu2012_PRC85-011301R,Lu2014_PRC89-014323,Zhao2015_PRC91-014321},
the ground state properties and PESs of the $^{270}$Hs
\cite{Meng2020_SciChinaPMA63-212011},
the nonaxial octupole $Y_{32}$ correlations in $N = 150$ isotones
\cite{Zhao2012_PRC86-057304}
and Zr isotopes \cite{Zhao2017_PRC95-014320},
octupole correlations in M$\chi$D of $^{78}$Br
\cite{Liu2016_PRL116-112501} and Ba isotopes \cite{Chen2016_PRC94-021301R}, etc.
In MDC-CDFTs, the reflection symmetry and the axial symmetry are both broken
and the shape degrees of freedom $\beta_{\lambda\mu}$ with $\mu$ being even numbers
are self-consistently included,
such as $\beta_{20},\beta_{22},\beta_{30},\beta_{32},\beta_{40},\beta_{42}$, and $\beta_{44}$.
Either the Bardeen-Cooper-Schrieffer (BCS) approach
or the Bogoliubov transformation
has been implemented to consider the pairing effects.
With two different ways of treating pairing correlations,
there are two types of MDC-CDFTs:
The one with the BCS approach is MDC relativistic mean-field (RMF) model,
the other with the Bogoliubov transformation is MDC relativistic Hartree-Bogoliubov (RHB) theory.

In this work, we use the MDC-RMF model to study
the ground state properties of SHN around
the doubly magic deformed nucleus $^{270}$Hs and
focus on the influence of the higher-order deformations.
This paper is organized as follows.
The MDC-CDFTs is introduced in Sec.~\ref{Sec:2}.
Then in Sec.~\ref{Sec:3} the results and discussions are presented.
Finally, we summarize this work in Sec.~\ref{Sec:4}.

\section{Theoretical framework}
\label{Sec:2}
In the CDFT, nucleons interact with each other through the exchange of
mesons and photons or point-coupling interaction.
To obtain correct saturation properties of nuclear matter,
the non-linear coupling terms or the density dependence of
the coupling constants are introduced.
Accordingly, there are four kinds of covariant density functionals:
The meson exchange (ME) or point-coupling (PC)
combined with the non-linear (NL) or density dependent (DD) couplings.
In this work, both the ME and PC density functionals are used.
The main formulae of the MDC-CDFTs can be found in
Refs.~\cite{Lu2014_PRC89-014323,Zhou2016_PS91-063008,
Zhao2017_PRC95-014320,Meng2020_SciChinaPMA63-212011}.
For convenience, here we only introduce the MDC-RMF
with the NL-PC effective interactions briefly.

The NL-PC Lagrangian is
\begin{equation}
	\mathcal{L}=\bar{\psi}\left(\mathrm{i} \gamma_{\mu} \partial^{\mu}
	- M\right) \psi
	- \mathcal{L}_{\mathrm{lin}}
	- \mathcal{L}_{\mathrm{nl}}
	- \mathcal{L}_{\mathrm{der}}
	- \mathcal{L}_{\mathrm{Cou}},
\end{equation}
where the linear, nonlinear, derivative couplings,
and the Coulomb terms are 
\begin{eqnarray}
	\mathcal{L}_{{\rm lin}} & = & \frac{1}{2} \alpha_{S} \rho_{S}^{2}
	+\frac{1}{2} \alpha_{V} \rho_{V}^{2}
	+\frac{1}{2} \alpha_{TS} \bm{\rho}_{TS}^{2} \nonumber \\
	& &+\frac{1}{2} \alpha_{TV} {\bm{\rho}_{TV}}^{2} ,
	\label{eq:L_lin}
	\\
	\mathcal{L}_{{\rm nl}}  & = & \frac{1}{3} \beta_{S} \rho_{S}^{3}
	+\frac{1}{4} \gamma_{S}\rho_{S}^{4}
	+\frac{1}{4} \gamma_{V}[\rho_{V}^{2}]^{2} ,
	\label{eq:L_nl}
	\\
	\mathcal{L}_{{\rm der}} & = & \frac{1}{2} \delta_{S}[\partial_{\nu}\rho_{S}]^{2}
	+\frac{1}{2} \delta_{V}[\partial_{\nu}\rho_{V}]^{2}
	+\frac{1}{2} \delta_{TS}[\partial_{\nu}\bm{\rho}_{TS}]^{2}
	\label{eq:L_der}
    \nonumber  \\
	&  & +\frac{1}{2} \delta_{TV}[\partial_{\nu}\bm{\rho}_{TV}]^{2} ,
	\\
	\mathcal{L}_{{\rm Cou}} & = & \frac{1}{4} F^{\mu\nu} F_{\mu\nu}
	+e\frac{1-\tau_{3}}{2} A_{0} \rho_{V} .
	\label{eq:L_Cou}
\end{eqnarray}
$M$ represents the nucleon mass and $e$ is the unit charge.
$\alpha_{S}$, $\alpha_{V}$, $\alpha_{TS}$, $\alpha_{TV}$, $\beta_{S}$,
$\gamma_{S}$, $\gamma_{V}$, $\delta_{S}$, $\delta_{V}$, $\delta_{TS}$, and $\delta_{TV}$
are coupling constants.
The isoscalar density $\rho_{S}$, isovector density $\bm{\rho}_{TS}$,
the time-like components of isoscalar current $\rho_{V}$, and
the time-like components of isovector currents $\bm{\rho}_{TV}$ are defined as
\begin{equation}
	\rho_{S}=\bar{\psi} \psi, \bm{\rho}_{T S}=\bar{\psi} \bm{\tau} \psi, \rho_{V}=\bar{\psi} \gamma^{0} \psi, \bm{\rho}_{TV}=\bar{\psi} \bm{\tau} \gamma^{0} \psi.
\end{equation}
The single particle wave function $\psi_{k}(\bm r)$ with the energy of $\epsilon_k$
of a nucleon is obtained by solving the Dirac equation
\begin{equation}
	\hat{h}\psi_{k}(\bm{r}) = \epsilon_{k} \psi_{k}(\bm{r}) ,
	\label{eq:Diracequation}
\end{equation}
with the Dirac Hamiltonian
\begin{equation}
	\hat{h} = \bm{\alpha} \cdot \bm{p}
	+ \beta \left[ M+S(\bm{r}) \right]
	+ V(\bm{r}),
	\label{eq:dirac}
\end{equation}
where the scalar potential $S(\bm{r})$ and vector potential $V(\bm{r})$ are
\begin{eqnarray}
\begin{aligned}
 S=&~ \alpha_{S} \rho_{S}
  +  \alpha_{T S} \bm{\rho}_{T S} \cdot \bm{\tau}
  +  \beta_{S} \rho_{S}^{2}
  +  \gamma_{S} \rho_{S}^{3} \\
 &+  \delta_{S} \Delta \rho_{S}
  +  \delta_{T S} \Delta \bm{\rho}_{T S} \cdot \bm{\tau}, \\
V=& ~  \alpha_{V} \rho_{V}
  +  \alpha_{T V} \bm{\rho}_{T V} \cdot \bm{\tau}
  +  \gamma_{V} \rho_{V}^{2} \rho_{V} \\
 &+  \delta_{V} \Delta \rho_{V}
  +  \delta_{T V} \Delta \bm{\rho}_{T V} \cdot \bm{\tau} + e\frac{1-\tau_{3}}{2}A_0.
\end{aligned}
\end{eqnarray}

In the MDC-CDFTs,
the wave functions are expanded in terms of
the axially deformed harmonic oscillator (ADHO) basis \cite{Gambhir1990_AP198-132,Ring1997_CPC105-77},
which is obtained by solving the Schr\"{o}dinger equation
\begin{eqnarray}
\left[-\frac{\hbar^{2}}{2M}\nabla^{2}+V_{B}(z,\rho)\right]\Phi_{\alpha}(\bm{r}\sigma)
	& = & E_{\alpha}\Phi_{\alpha}(\bm{r}\sigma),
	\label{eq:BasSchrodinger-1}
\end{eqnarray}
where $\bm{r} = (z,\rho)$ with $\rho=\sqrt{x^2+y^2}$
and
\begin{equation}
	V_{B}(z,\rho) = \frac{1}{2} M ( \omega_{\rho}^{2} \rho^{2} + \omega_{z}^{2} z^{2}),
\end{equation}
is the ADHO potential with the oscillator frequency
represented by $\omega_{\rho}$ ($\omega_{z}$)
perpendicular to (along) the $z$ axis.
More detailed formulae on the applications of ADHO in MDC-RMF can be find in
Refs. \cite{Lu2014_PRC89-014323,Zhou2016_PS91-063008,
Zhao2017_PRC95-014320,Meng2020_SciChinaPMA63-212011}.

After obtaining the ADHO basis, the single-particle wave functions can be expanded by using this basis
\begin{eqnarray}
	\psi_{i}(\bm{r}\sigma) =
	\left(
	\begin{array}{c}
		\sum_{\alpha}f_{i}^{\alpha} \Phi_{\alpha}(\bm{r}\sigma) \\
		\sum_{\alpha}g_{i}^{\alpha} \Phi_{\alpha}(\bm{r}\sigma)
	\end{array}
	\right),
	\label{eq:spwaveexpansion}
\end{eqnarray}
where $\alpha$ denotes a set of quantum numbers of the ADHO basis function,
$\alpha\equiv\{n_{z},n_{\rho},m_{l},m_{s}\}$.
$f_{i}^{\alpha}$ and
$g_{i}^{\alpha}$ are the expansion coefficients.
As for the truncation of the ADHO basis,
we follow Refs.~\cite{Warda2002_PRC66-014310,Gambhir1990_AP198-132}.
Finally, the wave functions are obtained by self-consistent iterations.

$^{270}$Hs is an axially deformed nuclei in the ground state
\cite{Meng2020_SciChinaPMA63-212011,Moller2016_ADNDT109--110-1,Jachimowicz2021_ADNDT138-101393}.
Reflection-asymmetric deformations occur normally only for ultra-neutron-rich nuclei
with $N \geqslant 182$ in SHN region
\cite{Jachimowicz2021_ADNDT138-101393}.
Thus we only consider the axially symmetric deformations
$\beta_\lambda$ with $\lambda$ being even numbers in the present work.
To study the influence of each shape degree of freedom on the bulk properties of SHN,
constraint calculations on mass multipole moments are performed \cite{Ring1980}.
In MDC-CDFTs, a modified linear-constraint method is implemented
\cite{Lu2012_PRC85-011301R,Lu2014_PRC89-014323}
and the Routhian reads
\begin{equation}
	E^{\prime} = E_{{\rm RMF}} +
	\sum_{\lambda} \frac{1}{2} C_{\lambda}Q_{\lambda} .
\end{equation}
After the $n$th iteration, the variable $C_{\lambda}^{(n+1)}$ is determined by
\begin{equation}
	C_{\lambda}^{(n+1)} =
	C_{\lambda}^{(n)} +
	k_{\lambda} \left( \beta_{\lambda}^{(n)} - \beta_{\lambda} \right),
\end{equation}
where $C_{\lambda}^{(n)}$ is the value in the $n$th iteration,
$k_{\lambda}$ is a constant and $\beta_{\lambda}$ is the
desired value of deformation parameter.

The intrinsic multipole moments are calculated as
\begin{equation}
	Q_{\lambda,\tau} = \int d^{3}\bm{r} \rho_{\tau}(\bm{r}) r^{\lambda} Y_{\lambda 0}(\Omega),
\end{equation}
where $\tau$ represents the nucleon, the neutron or the proton. $\rho_{\tau}$ is the corresponding vector density.
The deformation parameter $\beta_{\lambda,\tau}$ is given by
\begin{equation}
	\beta_{\lambda,\tau} = \frac{4\pi} {3N_{\tau}R^{\lambda}} Q_{\lambda,\tau},
\end{equation}
where $R= r_0 A^{{1}/{3}}$, the parameter $r_0=1.2$~fm,
and $N_\tau$ represents the corresponding particle's number $A$, $N$ or $Z$.

\section{Results and discussions}\label{Sec:3}

In order to study the influence of higher-order deformations
on the ground state properties of SHN,
the doubly magic deformed nucleus $^{270}$Hs,
even-even Hs isotopes from $^{264}$Hs to $^{276}$Hs
and $N=162$ isotones from $^{266}$Rf to $^{272}$Ds are analyzed.
When studying the ground state properties and
the influence of higher-order deformations,
the ADHO basis with $N_f=20$ shells is adopted,
leading to
an accuracy of 0.1 MeV in total energy of $^{270}$Hs
\cite{Meng2020_SciChinaPMA63-212011} in MDC-RMF caclulations.
In the particle-particle channel, a separable pairing force is adopted.
In this work, the strength and effective range of this force are taken to be the same as
those given in Ref. \cite{Meng2020_SciChinaPMA63-212011}:
$G=1.1 G_0$ with $G_0=728$ MeV fm$^3$ and $a=0.644$ fm.
The effective interactions
PC-PK1 \cite{Zhao2010_PRC82-054319},
PK1 \cite{Long2004_PRC69-034319},
PKDD \cite{Long2004_PRC69-034319},
DD-ME2 \cite{Lalazissis2005_PRC71-024312},
and NL3$^{*}$ \cite{Lalazissis2009_PLB671-36}
are used in the particle-hole channel.

\setlength{\LTcapwidth}{17 cm}
\setlength{\tabcolsep}{2.4mm}
\begin{longtable*}{ccccccccccccc}
	\caption{Ground state properties including
quadrupole deformation parameters of neutrons and protons ($\beta_{2,n}$ and $\beta_{2,p}$), deformation parameters $\beta_2$, $\beta_4$, $\beta_6$, $\beta_8$, and $\beta_{10}$, mass radius $R_t$,  radii of protons and neutrons ($R_p$ and $R_n$), charge radius $R_c$, and binding energy $E_\mathrm{B}$ of Hs isotopes by using MDC-RMF with five effective interactions PC-PK1, PK1, PKDD, DD-ME2, and NL3$^*$.} \label{tab:GS-1} \\
	\hline \hline
	~ & $\beta_{2,n}$ & $\beta_{2,p}$ & $\beta_{2}$ & $\beta_{4}$ & $\beta_{6}$ & $\beta_{8}$ & $\beta_{10}$ & $R_{n}$ & $R_{p}$ & $R_{\mathrm{t}}$ & $R_{\mathrm{c}}$  & $E_{\mathrm{B}}$ \\
	~ & ~ & ~ & ~ & ~ & ~ & ~ & ~ & (fm) & (fm) & (fm) & (fm)  & (MeV) \\
	\hline
	\endfirsthead
	\hline \hline
	~ & $\beta_{2,n}$ & $\beta_{2,p}$ & $\beta_{2}$ & $\beta_{4}$ & $\beta_{6}$ & $\beta_{8}$ & $\beta_{10}$ & $R_{n}$ & $R_{p}$ & $R_{\mathrm{t}}$ & $R_{\mathrm{c}}$  & $E_{\mathrm{B}}$ \\
	~ & ~ & ~ & ~ & ~ & ~ & ~ & ~ & (fm) & (fm) & (fm) & (fm)  & (MeV) \\
	\hline
	\endhead
	\hline \hline
	\endfoot
	PC-PK1 & ~ & ~ & ~ & ~ & ~ & ~ & ~ & ~ & ~ & ~ & ~ & \\
	$^{264}$Hs & 0.270 & 0.280 & 0.274 & $-$0.002 & $-$0.060 & $-$0.013 & 0.011 & 6.245 & 6.090 & 6.182 & 6.138 & 1924.415 \\
	$^{266}$Hs & 0.266 & 0.276 & 0.271 & $-$0.021 & $-$0.063 & $-$0.004 & 0.014 & 6.267 & 6.101 & 6.200 & 6.148 & 1939.205 \\
	$^{268}$Hs & 0.262 & 0.273 & 0.266 & $-$0.040 & $-$0.063 & ~~0.004  & 0.015 & 6.288 & 6.111 & 6.217 & 6.158 & 1953.554  \\
	$^{270}$Hs & 0.257 & 0.269 & 0.261 & $-$0.057 & $-$0.061 & ~~0.012 & 0.015 & 6.306 & 6.120 & 6.232 & 6.167 & 1967.408 \\
	$^{272}$Hs & 0.245 & 0.258 & 0.250 & $-$0.060 & $-$0.049 & ~~0.013 & 0.010 & 6.330 & 6.131 & 6.252 & 6.178 & 1979.303  \\
	$^{274}$Hs & 0.216 & 0.228 & 0.221 & $-$0.053 & $-$0.038 & ~~0.009 & 0.006 & 6.344 & 6.135 & 6.263 & 6.182 & 1990.951 \\
	$^{276}$Hs & 0.188 & 0.198 & 0.192 & $-$0.049 & $-$0.027 & ~~0.007 & 0.003 & 6.357 & 6.139 & 6.273 & 6.185 & 2002.778 \\
	PK1 & ~ & ~ & ~ & ~ & ~ & ~ & ~ & ~ & ~ & ~ & ~ & \\
	$^{264}$Hs & 0.253 & 0.258 & 0.255 & ~~0.006 & $-$0.058 & $-$0.016 & 0.011 & 6.228 & 6.058 & 6.159 & 6.105 & 1934.074 \\
	$^{266}$Hs & 0.253 & 0.258 & 0.255 & $-$0.014 & $-$0.065 & $-$0.006 & 0.016 & 6.253 & 6.070 & 6.179 & 6.118 & 1947.952 \\
	$^{268}$Hs & 0.256 & 0.261 & 0.258 & $-$0.034 & $-$0.070 & ~~0.005 & 0.019 & 6.278 & 6.084 & 6.201 & 6.131 & 1961.285  \\
	$^{270}$Hs & 0245 & 0.251 & 0.248 & $-$0.044 & $-$0.062 & ~~0.010 & 0.016 & 6.297 & 6.091 & 6.216 & 6.138 & 1973.766 \\
	$^{272}$Hs & 0.211 & 0.216 & 0.213 & $-$0.029 & $-$0.053 & ~~0.005 & 0.010 & 6.305 & 6.090 & 6.221 & 6.137 & 1985.924  \\
	$^{274}$Hs & 0.194 & 0.198 & 0.195 & $-$0.038 & $-$0.040 & ~~0.006 & 0.010 & 6.322 & 6.097 & 6.234 & 6.144 & 1997.412 \\
	$^{276}$Hs & 0.178 & 0.182 & 0.180 & $-$0.047 & $-$0.028 & ~~0.007 & 0.007 & 6.342 & 6.105 & 6.250 & 6.151 & 2008.356 \\
	PKDD & ~ & ~ & ~ & ~ & ~ & ~ & ~ & ~ & ~ & ~ & ~ & \\
	$^{264}$Hs & 0.250 & 0.255 & 0.252 & ~~0.001 & $-$0.060 & $-$0.015 & 0.011 & 6.207 & 6.053 & 6.145 & 6.101 & 1932.544 \\
	$^{266}$Hs & 0.253 & 0.258 & 0.255 & $-$0.020 & $-$0.066 & $-$0.004 & 0.016 & 6.233 & 6.067 & 6.166 & 6.115 & 1946.294 \\
	$^{268}$Hs & 0.258 & 0.264 & 0.260 & $-$0.041 & $-$0.072 & ~~0.009 & 0.021 & 6.259 & 6.082 & 6.188 & 6.129 & 1959.686  \\
	$^{270}$Hs & 0.252 & 0.261 & 0.256 & $-$0.059 & $-$0.062 & ~~0.017 & 0.017 & 6.278 & 6.091 & 6.204 & 6.138 & 1972.399  \\
	$^{272}$Hs & 0.211 & 0.217 & 0.213 & $-$0.030 & $-$0.056 & ~~0.006 & 0.019 & 6.282 & 6.087 & 6.205 & 6.134 & 1983.376  \\
	$^{274}$Hs & 0.190 & 0.194 & 0.191 & $-$0.039 & $-$0.041 & ~~0.006 & 0.010 & 6.296 & 6.093 & 6.216 & 6.139 & 1994.504 \\
	$^{276}$Hs & 0.174 & 0.179 & 0.176 & $-$0.048 & $-$0.028 & ~~0.007 & 0.007 & 6.316 & 6.100 & 6.233 & 6.147 & 2004.934 \\
	DD-ME2  & ~ & ~ & ~ & ~ & ~ & ~ & ~ & ~ & ~ & ~ & ~ & \\
	$^{264}$Hs & 0.260 & 0.267 & 0.263 & $-$0.001 & $-$0.061 & $-$0.012 & 0.014 & 6.178 & 6.073 & 6.136 & 6.121 & 1928.426 \\
	$^{266}$Hs & 0.261 & 0.269 & 0.264 & $-$0.023 & $-$0.066 & $-$0.001 & 0.019 & 6.200 & 6.086 & 6.154 & 6.133 & 1942.991 \\
	$^{268}$Hs & 0.259 & 0.269 & 0.263 & $-$0.042 & $-$0.068 & ~~0.011 & 0.021 & 6.220 & 6.097 & 6.171 & 6.144 & 1957.270  \\
	$^{270}$Hs & 0.252 & 0.264 & 0.257 & $-$0.058 & $-$0.060 & ~~0.017 & 0.017 & 6.236 & 6.105 & 6.184 & 6.152 & 1971.027  \\
	$^{272}$Hs & 0.213 & 0.222 & 0.216 & $-$0.032 & $-$0.054 & ~~0.007 & 0.012 & 6.240 & 6.101 & 6.185 & 6.148 & 1981.994  \\
	$^{274}$Hs & 0.196 & 0.204 & 0.199 & $-$0.039 & $-$0.041 & ~~0.007 & 0.010 & 6.256 & 6.107 & 6.198 & 6.154 & 1993.391 \\
	$^{276}$Hs & 0.178 & 0.186 & 0.181 & $-$0.048 & $-$0.027 & ~~0.008 & 0.007 & 6.271 & 6.113 & 6.210 & 6.160 & 2004.657 \\
	NL3$^*$ & ~ & ~ & ~ & ~ & ~ & ~ & ~ & ~ & ~ & ~ & ~ & \\
	$^{264}$Hs & 0.265 & 0.271 & 0.267 & ~~0.004 & $-$0.060 & $-$0.014 & 0.013 & 6.260 & 6.079 & 6.186 & 6.126 & 1931.827 \\
	$^{266}$Hs & 0.263 & 0.270 & 0.266 & $-$0.017 & $-$0.065 & $-$0.004 & 0.017 & 6.284 & 6.090 & 6.206 & 6.137 & 1945.933 \\
	$^{268}$Hs & 0.262 & 0.269 & 0.265 & $-$0.036 & $-$0.067 & ~~0.007 & 0.018 & 6.307 & 6.101 & 6.225 & 6.148 & 1959.582  \\
	$^{270}$Hs & 0.256 & 0.265 & 0.260 & $-$0.054 & $-$0.061 & ~~0.015 & 0.017 & 6.326 & 6.109 & 6.240 & 6.156 & 1972.574  \\
	$^{272}$Hs & 0.233 & 0.241 & 0.236 & $-$0.045 & $-$0.051 & ~~0.010 & 0.012 & 6.344 & 6.115 & 6.254 & 6.162 & 1984.027  \\
	$^{274}$Hs & 0.204 & 0.211 & 0.207 & $-$0.039 & $-$0.040 & ~~0.007 & 0.009 & 6.357 & 6.118 & 6.264 & 6.164 & 1995.465 \\
	$^{276}$Hs & 0.185 & 0.193 & 0.188 & $-$0.047 & $-$0.028 & ~~0.008 & 0.006 & 6.376 & 6.125 & 6.279 & 6.171 & 2006.648 \\
\end{longtable*}

\setlength{\tabcolsep}{2.4mm}
\begin{longtable*}{ccccccccccccc}
	\caption{Same as Table \ref{tab:GS-1}, but for isotones with $N=162$.} \label{tab:GS-2} \\
	\hline \hline
	~ & $\beta_{2,n}$ & $\beta_{2,p}$ & $\beta_{2}$ & $\beta_{4}$ & $\beta_{6}$ & $\beta_{8}$ & $\beta_{10}$ & $R_{n}$ & $R_{p}$ & $R_{\mathrm{t}}$ & $R_{\mathrm{c}}$  & $E_{\mathrm{B}}$ \\
	~ & ~ & ~ & ~ & ~ & ~ & ~ & ~ & (fm) & (fm) & (fm) & (fm)  & (MeV) \\
	\hline
	\endfirsthead
	\hline \hline
	~ & $\beta_{2,n}$ & $\beta_{2,p}$ & $\beta_{2}$ & $\beta_{4}$ & $\beta_{6}$ & $\beta_{8}$ & $\beta_{10}$ & $R_{n}$ & $R_{p}$ & $R_{\mathrm{t}}$ & $R_{\mathrm{c}}$  & $E_{\mathrm{B}}$ \\
	~ & ~ & ~ & ~ & ~ & ~ & ~ & ~ & (fm) & (fm) & (fm) & (fm)  & (MeV) \\
	\hline
	\endhead
	\hline \hline
	\endfoot
	PC-PK1 & ~ & ~ & ~ & ~ & ~ & ~ & ~ & ~ & ~ & ~ & ~ & \\
	$^{266}$Rf & 0.261 & 0.274 & 0.266 & $-$0.039 & $-$0.060 & 0.005 & 0.013 & 6.303 & 6.083 & 6.218 & 6.129 & 1953.531  \\
	$^{268}$Sg & 0.260 & 0.274 & 0.266 & $-$0.048 & $-$0.063 & 0.009 & 0.015 & 6.304 & 6.102 & 6.225 & 6.149 & 1961.448  \\
	$^{270}$Hs & 0.257 & 0.269 & 0.261 & $-$0.057 & $-$0.061 & 0.012 & 0.015 & 6.306 & 6.120 & 6.232 & 6.167 & 1967.408 \\
	$^{272}$Ds & 0.249 & 0.258 & 0.253 & $-$0.061 & $-$0.055 & 0.015 & 0.012 & 6.308 & 6.139 & 6.240 & 6.186 & 1971.338  \\
	PK1 & ~ & ~ & ~ & ~ & ~ & ~ & ~ & ~ & ~ & ~ & ~ & \\
	$^{266}$Rf & 0.256 & 0.265 & 0.260 & $-$0.038 & $-$0.062 & 0.007 & 0.017 & 6.296 & 6.053 & 6.202 & 6.100 & 1957.699  \\
	$^{268}$Sg & 0.257 & 0.268 & 0.261 & $-$0.046 & $-$0.067 & 0.010 & 0.018 & 6.299 & 6.075 & 6.211 & 6.122 & 1966.814  \\
	$^{270}$Hs & 0.245 & 0.251 & 0.248 & $-$0.044 & $-$0.062 & 0.010 & 0.016 & 6.297 & 6.091 & 6.216 & 6.138 & 1973.766 \\
	$^{272}$Ds & 0.225 & 0.228 & 0.226 & $-$0.037 & $-$0.054 & 0.007 & 0.012 & 6.292 & 6.104 & 6.217 & 6.151 & 1979.041  \\
	PKDD & ~ & ~ & ~ & ~ & ~ & ~ & ~ & ~ & ~ & ~ & ~ & \\
	$^{266}$Rf & 0.256 & 0.266 & 0.260 & $-$0.043 & $-$0.063 & 0.009 & 0.018 & 6.272 & 6.050 & 6.186 & 6.097 & 1955.230  \\
	$^{268}$Sg & 0.259 & 0.272 & 0.264 & $-$0.051 & $-$0.070 & 0.013 & 0.020 & 6.277 & 6.073 & 6.197 & 6.120 & 1965.167  \\
	$^{270}$Hs & 0.252 & 0.261 & 0.256 & $-$0.059 & $-$0.062 & 0.017 & 0.017 & 6.278 & 6.091 & 6.204 & 6.138 & 1972.399  \\
	$^{272}$Ds & 0.241 & 0.245 & 0.242 & $-$0.060 & $-$0.056 & 0.017 & 0.015 & 6.278 & 6.107 & 6.210 & 6.154 & 1977.775  \\
	DD-ME2 & ~ & ~ & ~ & ~ & ~ & ~ & ~ & ~ & ~ & ~ & ~ & \\
	$^{266}$Rf & 0.255 & 0.268 & 0.260 & $-$0.040 & $-$0.061 & 0.008 & 0.017 & 6.224 & 6.062 & 6.161 & 6.109 & 1955.711  \\
	$^{268}$Sg & 0.257 & 0.271 & 0.262 & $-$0.048 & $-$0.065 & 0.012 & 0.019 & 6.231 & 6.086 & 6.174 & 6.132 & 1964.571  \\
	$^{270}$Hs & 0.252 & 0.264 & 0.257 & $-$0.058 & $-$0.060 & 0.017 & 0.017 & 6.236 & 6.105 & 6.184 & 6.152 & 1971.027  \\
	$^{272}$Ds & 0.242 & 0.249 & 0.245 & $-$0.060 & $-$0.054 & 0.017 & 0.015 & 6.241 & 6.123 & 6.193 & 6.170 & 1975.320  \\
	NL3$^*$ & ~ & ~ & ~ & ~ & ~ & ~ & ~ & ~ & ~ & ~ & ~ & \\
	$^{266}$Rf & 0.261 & 0.272 & 0.266 & $-$0.038 & $-$0.062 & 0.007 & 0.017 & 6.325 & 6.070 & 6.226 & 6.116 & 1956.503  \\
	$^{268}$Sg & 0.261 & 0.273 & 0.266 & $-$0.047 & $-$0.066 & 0.011 & 0.018 & 6.325 & 6.091 & 6.233 & 6.137 & 1965.582  \\
	$^{270}$Hs & 0.256 & 0.265 & 0.260 & $-$0.054 & $-$0.061 & 0.015 & 0.017 & 6.326 & 6.109 & 6.240 & 6.156 & 1972.574  \\
	$^{272}$Ds & 0.248 & 0.253 & 0.250 & $-$0.058 & $-$0.055 & 0.016 & 0.014 & 6.326 & 6.127 & 6.247 & 6.174 & 1977.689  \\
\end{longtable*}

\begin{figure*}[htb]
	\begin{center}
		\includegraphics[width=0.9\textwidth]{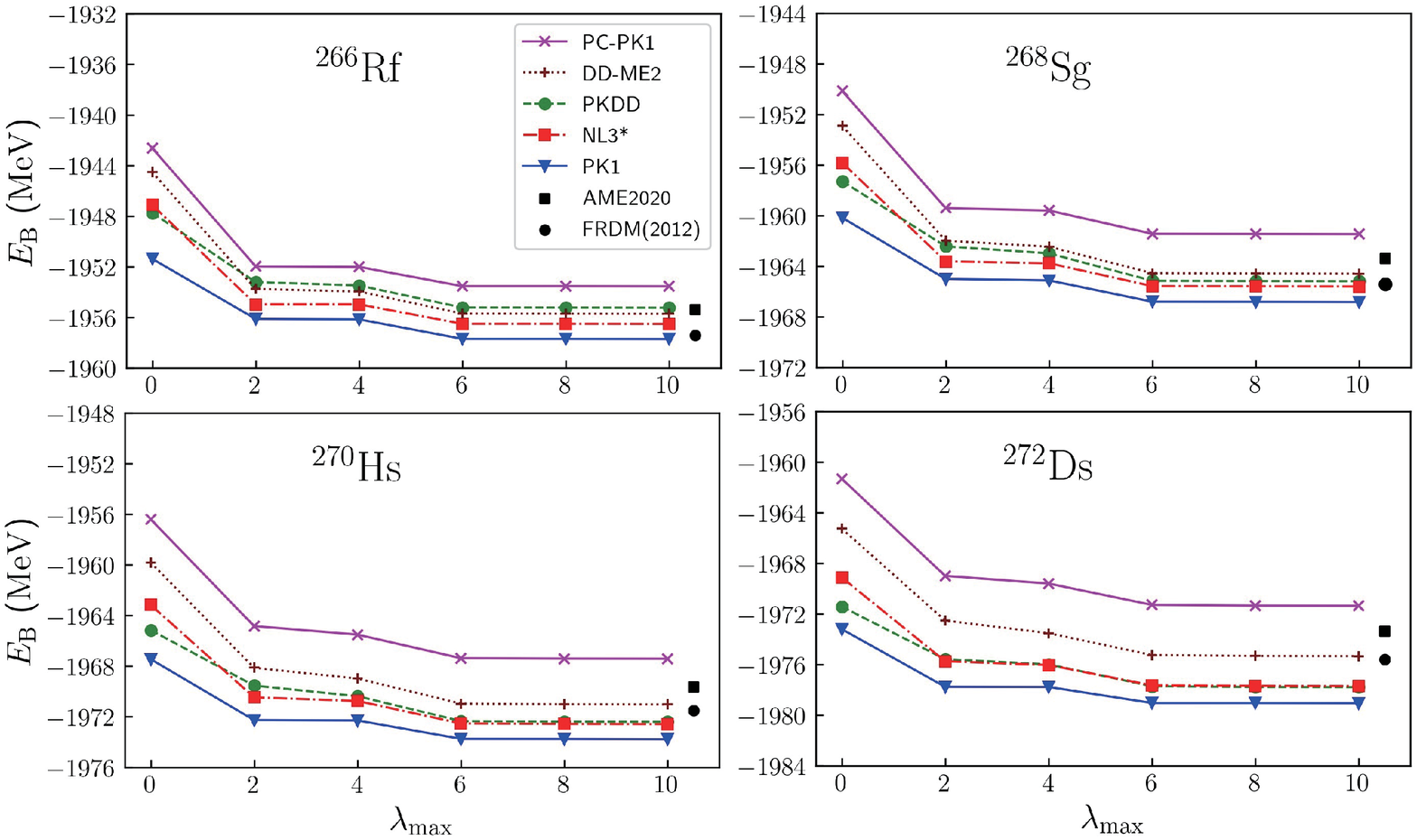}
	\end{center}
	\caption{Binding energies of isotones with $N=162$ with PC-PK1, DD-ME2, PKDD, NL3$^{*}$, and PK1
     as a function of $\lambda_\mathrm{max}$.
     The black square and black point represent the value of AME2020 and FRMD(2012), respectively.}
	\label{fig:BE-isotopes}
\end{figure*}

\begin{figure*}[htb]
	\begin{center}
		\includegraphics[width=0.9\textwidth]{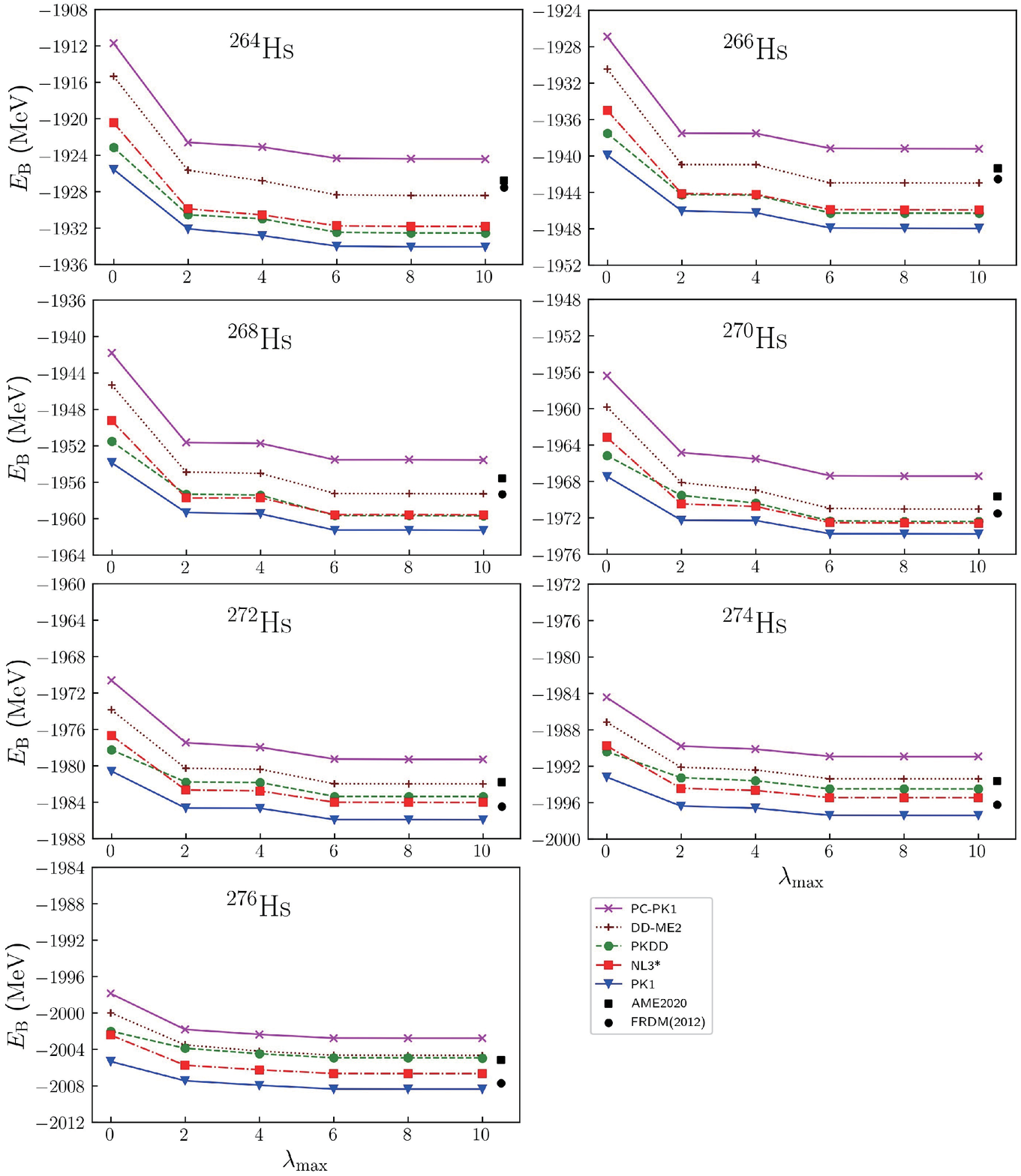}
	\end{center}
	\caption{Binding energies of Hs isotopes with PC-PK1, DD-ME2, PKDD, NL3$^{*}$, and PK1
     as a function of $\lambda_\mathrm{max}$.
     The black square and black point represent the value of AME2020 and FRMD(2012), respectively.}
	\label{fig:BE-isotones}
\end{figure*}

\begin{figure*}[htb]
	\centering
	\includegraphics[width=\textwidth]{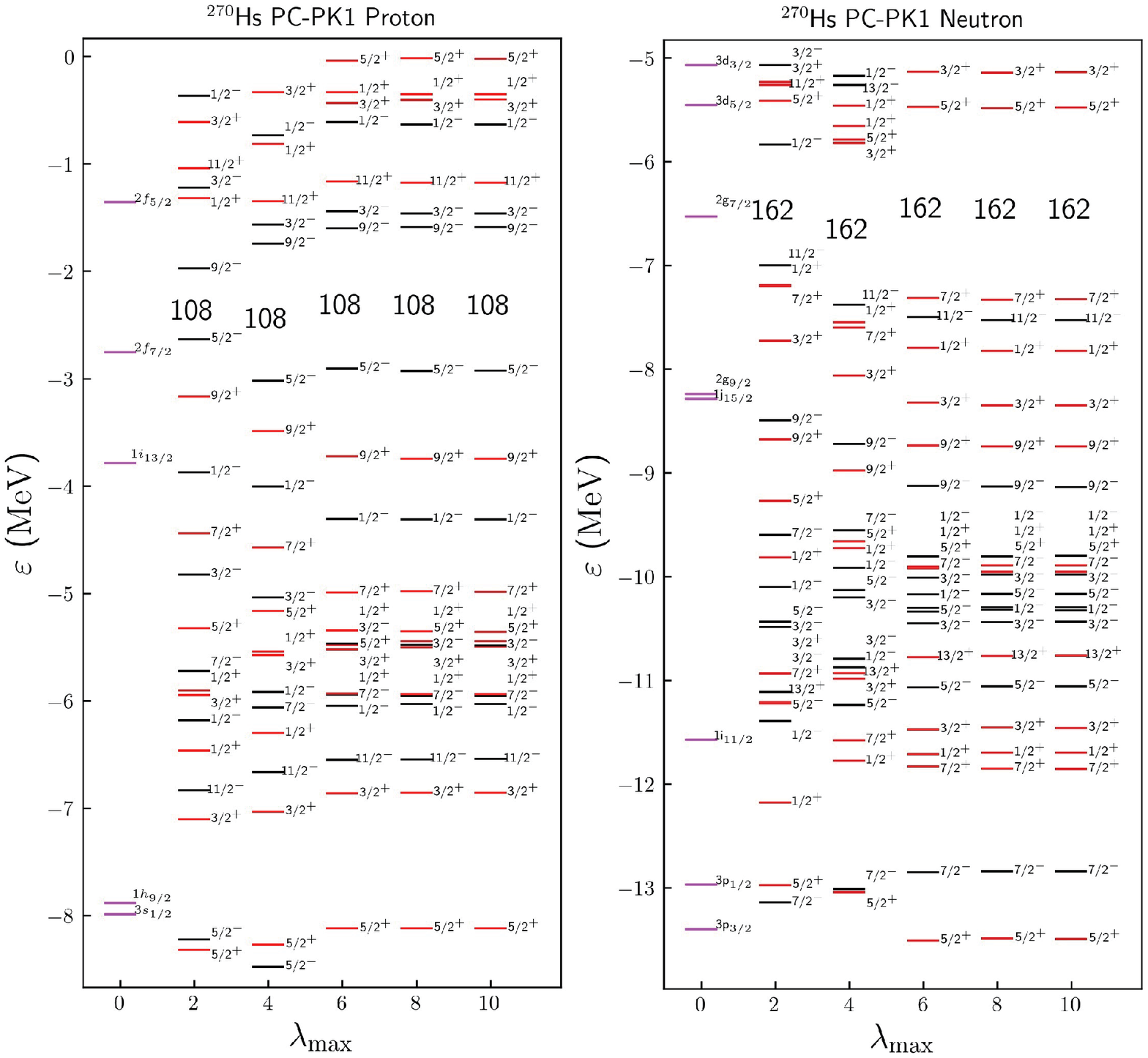}
	\caption{Single proton and neutron levels of $^{270}$Hs from PC-PK1 calculations with different $\lambda_\mathrm{max}$. In the spherical case ($\lambda_\mathrm{max}=0$), each level is labelled by $|nlj\rangle$.
When $\lambda_\mathrm{max}\neq0$, each level is labelled by the projection $\Omega$ of total angular momentum on the symmetry axis and the parity $\pi$.
Single-particle levels with positive and negative parities are presented by red and black lines, respectively.}
	\label{fig:PN}
\end{figure*}

The ground state properties, including deformation parameters $\beta_\lambda$ ($\lambda = 2, 4, 6, 8$, and $10$),
radii, and binding energies of even-even Hs isotopes
with above-mentioned five effective interactions
are given in Table \ref{tab:GS-1} and
even-even isotones with $N=162$ are listed in Table \ref{tab:GS-2}.
One could find that the binding energies of one nucleus with five effective interactions
differ from each other, e.g.,
the largest binding energy of $^{270}$Hs is 1973.77 MeV with PK1
and the smallest is 1967.41 MeV with PC-PK1.
Such results are relatively close to the empirical value in AME2020
$E_{\mathrm{B}} = 1969.65$ MeV \cite{Kondev2021_ChinPC45-030001,Huang2021_ChinPC45-030002,Wang2021_ChinPC45-030003}
and also comparable to the prediction of other models,
such as MMM $E_{\mathrm{B}} = 1969.20$ MeV \cite{Patyk1991_NPA533-132},
the Skyrme Hartree-Fock Bogoliubov mass model (HFB-24) $E_{\mathrm{B}} = 1968.45$ MeV \cite{Goriely2013_PRC88-024308},
the Weizs\"{a}cker-Skyrme (WS) mass formula WS4
 $E_{\mathrm{B}} = 1970.27$ MeV \cite{Wang2014_PLB734-215},
the finite range droplet model (FRDM(2012))
 $E_{\mathrm{B}} = 1971.48$ MeV \cite{Moller2016_ADNDT109--110-1},
and several RMF calculations
\cite{Ren2002_PRC65-051304R,Ren2002_PRC66-064306,
Geng2005_PTP113-785,
Zhang2012_PRC85-014325,Shi2019_ChinPC43-074104}.
For other nuclei, similar conclusions can also be obtained.
From these two tables, it is obvious that all the nuclei involved in the present work
are deformed in MDC-RMF calculations with five effective interactions.
This is consistent with the results shown in MMM calculations
\cite{Patyk1991_NPA533-132,Moller2016_ADNDT109--110-1,
Jachimowicz2021_ADNDT138-101393} and the other global studies
\cite{Agbemava2014_PRC89-054320,Erler2012_Nature486-509}.
In addition,
it has been shown that the inclusion of the rotational energy correction (REC)
can improve the description of binding energies with PC-PK1
\cite{Zhao2010_PRC82-054319}.
In this work, after considering RECs in PC-PK1 calculations,
the binding energy of $^{270}$Hs changes from $1967.45$ MeV to $1969.76$ MeV,
which is more close to the value given in AME2020.

To determine the dimension of the deformation space
when studying the ground states of SHN by using MDC-CDFTs,
we calculate the binding energies of Hs isotopes and isotones with $N=162$
in different deformation space $\{\beta_\lambda;\lambda=0,2,\cdots, \lambda_\mathrm{max}\}$ with
$\lambda_\mathrm{max}$ being the maximum order of deformation parameters,
which means that all the deformation parameters
$\beta_{\lambda} \leq \beta_{\lambda_\mathrm{max}}$ are considered self-consistently
while other deformation parameters are constrained to be zero.
In Fig.~\ref{fig:BE-isotopes}, the binding energies of $N=162$ isotones
with five different effective interactions
are plotted as function of $\lambda_{\mathrm{max}}$.
For convenience, here we take $^{270}$Hs with the effective interaction PC-PK1
as an example to discuss the influence of each order of deformation on the binding energy
in detail.
When constraining $^{270}$Hs to be spherical,
i.e., in the deformation space $\left\{\beta_\lambda; \lambda=0\right\}$,
the resulting binding energy is 1956.39 MeV,
which is close to the prediction of relativistic continuum Hartree-Bogoliubov theory
1952.65 MeV \cite{Xia2018_ADNDT121--122-1}
but far from the value given in AME2020
(marked by black square in Fig.~\ref{fig:BE-isotopes}).
After taking the quadrupole deformation $\beta_2$ into account,
the binding energy of $^{270}$Hs changes very much (about 8.43 MeV)
and becomes closer to that in AME2020.
This result indicates the importance of the quadrupole deformation.
The influence of the hexadecapole deformation $\beta_4$
on the binding energy is not so big,
only 0.68 MeV.
If we further consider $\beta_6$,
the change of energy is about 1.87 MeV,
which is much larger than that corresponding to $\beta_4$
and $E_\mathrm{B}$ approaches to the value given in AME2020.
Including $\beta_8$ and $\beta_{10}$
almost does not affect the binding energy,
which converges well at $\{\beta_\lambda;\lambda=0,2,\cdots,10\}$.
From these results, we can conclude that to get a proper description of $^{270}$Hs,
one should consider the $\beta_6$ deformation at least
from the point of view of binding energy.
Calculated binding energy versus $\lambda_{\mathrm{max}}$ with
other density functionals are also shown in Fig.~\ref{fig:BE-isotopes} and
one can find that although the binding energies with
five effective interactions differ from each other,
the overall trends that $E_\mathrm{B}$ changes with $\lambda_{\mathrm{max}}$ are similar.
The binding energy of $^{270}$Hs is largely changed by $\beta_2$,
then $\beta_6$ and $\beta_4$.
The influence of $\beta_8$ and $\beta_{10}$ can be ignored.
As for the RECs for $^{270}$Hs with PC-PK1,
they are 2.27, 2.03, 2.29, 2.31, and 2.31 MeV in
deformation spaces $\{\beta_\lambda;\lambda=0,\cdots, \lambda_\mathrm{max}\}$ with
$\lambda_{\mathrm{max}}=2,4,6,8$, and $10$,
respectively. The values of RECs change slightly in different deformation space and almost do not
influence the trends of binding energies with respect to $\lambda_\mathrm{max}$.

To check whether the conclusion mentioned-above is valid for other SHN,
we performed similar calculations for even-even isotones with $N=162$ and Hs isotopes
and results are also presented in Fig.~\ref{fig:BE-isotopes} and \ref{fig:BE-isotones}.
From these figures, we can find that the binding energies of these nuclei are
significantly changed by $\beta_2$.
The influence of $\beta_4$ and $\beta_6$ cannot be ignored and
the contribution to total energy from $\beta_6$
is larger than that from $\beta_4$.
For Hs isotopes, with the decease of the neutron number,
the value of $\beta_2$ increases a lot and the differences of total energies between in the spherical case and in ground states become larger,
which can be seen from Fig. \ref{fig:BE-isotones}.
For isotones with $N=162$, the value of $\beta_2$ changes not so big
with the proton number.
So for these nuclei, the trends of binding energies
with respect to $\beta_{\lambda_\mathrm{max}}$
keep the same as that for $^{270}$Hs.

It is well known that the shell structure is especially important for SHN
and very sensitive to the deformation of the nucleus
\cite{Patyk1991_PLB256-307}.
We take $^{270}$Hs as an example again to explore how the deformations influence the shell gaps
at $Z = 108$ and $N = 162$
by studying the structure of single-particle levels (SPLs) in different deformation spaces.
In Fig.~\ref{fig:PN} we show the SPLs for protons and
neutrons of $^{270}$Hs versus $\lambda_\mathrm{max}$, calculated with PC-PK1.
When $\lambda_{\mathrm{max}} = 10$, i.e., for the ground state, the
energy gaps at $Z = 108$ and $N = 162$ are about 1.34 MeV and 1.85 MeV,
which are considerably large for such a heavy nucleus
\cite{Agbemava2015_PRC92-054310} and result in deformed shells.

In the spherical limit, $\lambda_{\mathrm{max}}=0$,
each single particle state is labelled by $|nlj\rangle$
where $n$, $l$, and $j$ denote the radial quantum number,
orbital angular momentum, and total angular momentum, respectively.
It is obvious that there are no shell gaps at $Z=108$ and $N=162$.
After including $\beta_2$,
a spherical orbital $|nlj\rangle$  with the degeneracy of $2j+1$
splits into $(2j+1)/2$
levels and each one is represented by $\Omega^\pi$ with
the projection $\Omega$ of total angular momentum on the symmetry axis
and the parity $\pi$.
It is found that due to quadrupole correlations
the shell gaps at $Z=108$ and $N=162$ appear,
0.66 MeV and 1.17 MeV, respectively.
When including the $\beta_4$ into the deformation space,
the order of SPLs around two gaps changes and the shell gaps at $Z=108$
(up to about 1.28 MeV) and $N=162$ (up to about 1.56 MeV)
increase largely.
The impact of $\beta_6$ on the shell gap at $Z=108$ is not so big,
only 0.02 MeV, but for neutrons the shell gap at $N=162$ increases about 0.26 MeV.
The inclusion of $\beta_8$ and $\beta_{10}$ almost does not change the shell gaps
and the order of SPLs.
From these discussions, one can conclude that
$\beta_2$ plays a vital role for the formation of the shell closures $Z=108$ and $N=162$,
which are further enhanced by $\beta_4$.
The influence of $\beta_6$ is relatively small and
the effects of $\beta_8$ and $\beta_{10}$ can be negligible.
There remains a question: Where do the $Y_{60}$ correlations come from?
By checking the SPLs, we find that two proton levels $1/2^+$ originating from
the spherical orbitals $3s_{1/2}$ and $1i_{13/2}$ are very close to
each other and the mixing of these two spherical orbital in the deformed SPLs results in $Y_{60}$ correlations.
For neutrons,
these correlations originate from the mixing of
the spherical orbitals 3$p_{3/2}$ and 1$j_{15/2}$
in the levels $3/2^-$ close to the neutron Fermi energy.

\section{Summary}
\label{Sec:4}
In this work, we investigate the ground state properties of SHN around $^{270}$Hs
in multidimensional deformation spaces by
using the MDC-RMF model with five density functionals.
The influence of higher-order deformation parameters on the ground state of nuclei near $^{270}$Hs are studied,
including the binding energies and SPLs.
We have shown that the binding energies of deformed SHN around $^{270}$Hs are
significantly affected by the higher-order deformations.
In particular, the influence of $\beta_6$ on binding energy is larger than that from $\beta_4$.
For doubly magic nucleus $^{270}$Hs,
the deformed shell gaps at $Z=108$ and $N=162$ are mainly determined by quarupole correlations
and enhanced by the inclusion of $\beta_4$.
In conclusion, the $\beta_6$ degree of freedom should be considered at least in the study of SHN by using CDFTs.
It is also very interesting to study how the higher-order deformations influence other properties of SHN, such as moment
of inertia, energy spectra by using density functional theories.
In addition, we would like to mention that
the calculations performed in this work can also be done with the deformation relativistic Hartree-Bogoliubov (DRHBc) theory
\cite{Zhou2010_PRC82-011301R,Li2012_PRC85-024312,Sun2018_PLB785-530,
Sun2020_NPA1003-122011,Sun2021_PRC103-054315,Sun2021_SciBulletin66-2072,Sun2021_arXiv2107.05925}, in which the scalar potential and densities are
expanded in terms of the Legendre polynomials
but the time consuming of the DRHBc theory is much heavier than that for MDC-RMF.
Very recently, the influence of higher-order deformation on possible bound nuclei beyond the
drip line has been investigated
in the transfermium region from No ($Z=102$) to Ds ($Z=110$)
by using the DRHBc theory \cite{He2021_ChinPC45-101001}
and a nuclear mass table with the DRHBc theory is in progress
\cite{Zhang2020_PRC102-024314,In2021_IJMPE30-2150009,Zhang2021_PRC104-L021301,
Pan2021_PRC104-024331}.

\acknowledgments
We thank Bin-Nan Lu, Yu-Ting Rong, and Kun Wang for helpful discussions.
This work has been supporteds by
the National Key R\&D Program of China (Grant No. 2018YFA0404402),
the National Natural Science Foundation of China (Grants
No. 11525524, No. 12070131001, No. 12047503, No.11975237, and No. 11961141004),
the Key Research Program of Frontier Sciences of Chinese Academy of Sciences
(Grant No. QYZDB-SSWSYS013),
the Strategic Priority Research Program of Chinese Academy of Sciences
(Grant No. XDB34010000 and No. XDPB15),
and the IAEA Coordinated Research Project (Grant No. F41033).
The results described in this paper are obtained on
the High-performance Computing Cluster of ITP-CAS and
the ScGrid of the Supercomputing Center,
Computer Network Information Center of Chinese Academy of Sciences.


%

\end{document}